\documentclass[prl,aps,superscriptaddress,twocolumn]{revtex4-2}

\usepackage{bm,bbm,color,float,dcolumn,amsmath,amssymb,graphicx,subfigure}
\usepackage{physics}
\usepackage[colorlinks=true]{hyperref}
\hypersetup{linkcolor=blue,citecolor=blue,urlcolor=blue}

\definecolor{psychedelicpurple}{rgb}{0.87, 0.0, 1.0}
\definecolor{violet}{rgb}{0.56, 0, 1}

\begin{document}

\title{Finite-size corrections to the crosscap overlap in the two-dimensional Ising model}

\author{Yiteng Zhang}
\email{z.yt-wangzai@connect.hku.hk}
\affiliation{Department of Physics, The University of Hong
Kong, Pokfulam Road, Hong Kong, China}

\author{Li-Ping Yang}
\email{liping2012@cqu.edu.cn}
\affiliation{Department of Physics and Chongqing Key Laboratory for Strongly Coupled Physics, Chongqing University, Chongqing 401331, China}

\author{Hong-Hao Tu}
\email{h.tu@lmu.de}
\affiliation{Faculty of Physics and Arnold Sommerfeld Center for Theoretical Physics, Ludwig-Maximilians-Universit\"at M\"unchen, 80333 Munich, Germany}

\author{Yueshui Zhang}
\email{yueshui@hku.hk}
\affiliation{Department of Physics, The University of Hong
Kong, Pokfulam Road, Hong Kong, China}
\affiliation{Faculty of Physics and Arnold Sommerfeld Center for Theoretical Physics, Ludwig-Maximilians-Universit\"at M\"unchen, 80333 Munich, Germany}

\date{\today}

\begin{abstract}
We analyze the finite-size corrections to the crosscap overlap in the two-dimensional classical Ising model along its self-dual critical line. Using a fermionic formulation, we express the lattice crosscap overlap in terms of Bogoliubov angles and develop a contour-integral approach by analytically continuing the lattice momentum to the complex plane. This leads to a remarkably simple expression for the crosscap overlap, which demonstrates that the finite-size corrections decay exponentially with system size. We further derive an exact analytical formula for the corresponding decay constant and show that it is determined by the complex singularity structure of the Bogoliubov angle.
\end{abstract}

\maketitle

{\em Introduction} --- Boundary effects are fundamental to the understanding of critical phenomena in low-dimensional systems. At critical points, the low-energy physics of many systems is described by conformal field theories (CFTs), for which conformal boundary conditions give rise to universal quantities characterizing the underlying critical theory~\cite{Cardy1984,Cardy1986b,Cardy1989,Behrend1998,Behrend2000}. A paradigmatic example is the boundary entropy introduced by Affleck and Ludwig~\cite{Affleck1991a,Friedan2004}, which quantifies the effective ground-state degeneracy associated with a conformal boundary and serves as an important characteristic quantity for boundary critical phenomena.

In two dimensions, boundary critical phenomena are far more analytically tractable due to the infinite-dimensional conformal symmetry of the underlying CFT. In the path integral formulation, a boundary of the two-dimensional CFT corresponds geometrically to an edge of the space‑time manifold, forming a circle, to which an associated boundary condition is assigned. Alongside conventional boundaries, nonorientable structures provide additional probes of CFT data. A crosscap, constructed by identifying antipodal points of the circle, serves as a fundamental building block for defining CFTs on nonorientable manifolds, such as the Klein bottle and the real projective plane~\cite{Ishibashi1989,Fioravanti1994}. When the circle, equipped with the crosscap boundary condition, is taken as the spatial direction, the quantization procedure yields the crosscap state, and the crosscap overlap is defined as its overlap with the CFT vacuum~\cite{Fioravanti1994,Pradisi1995,Blumenhagen-Book,ZhangYS2023}. The logarithm of the squared norm of this overlap, known as the Klein bottle entropy~\cite{TuHH2017}, is a universal quantity that encodes essential CFT data and can be regarded as the non‑orientable analog of the Affleck‑Ludwig boundary entropy in conventional boundary CFTs. The study of crosscap states and the associated Klein bottle entropy has seen considerable recent development across condensed matter and statistical physics, as well as high-energy theory and mathematical physics, with these concepts extended both at and near conformal critical points and applied to various one-dimensional (1D) quantum and two-dimensional (2D) classical models, in both equilibrium and nonequilibrium contexts~\cite{TangW2017,ChenL2017,WangHX2018,TangW2019,LiZQ2020,Garcia2021,Caetano2022,Gombor2022,Ekman2022,Vanhove2022,HeM2023,Cimasoni2024,Shimizu2026,TanBY2025,ZhangYS2026a,WeiZX2025,LiYZ2025,Harada2026,Yoneta2024,Chalas2025a,WeiZX2026,ChenHH2026,BaiC2026,Dulac2026,Chalas2025b}. An important ongoing effort is to construct lattice crosscap states in concrete models and to establish their identification with the corresponding CFT counterparts~\cite{TanBY2025,ZhangYS2026a}.

It is useful to contrast the crosscap boundary condition with more familiar boundary choices. Periodic and antiperiodic boundary conditions preserve orientability and define, respectively, the ordinary and $\mathbb{Z}_2$-twisted torus partition functions. Their finite-size corrections are typically organized as algebraic expansions in the inverse system size (see, e.g., Ref.~\cite{Izmailian2001}). Open (free) and fixed boundaries, by contrast, are described by conformal boundary states~\cite{Cardy1986b,Cardy1989}. Their free energies contain nonuniversal surface and, when applicable, corner contributions, together with the universal Affleck-Ludwig boundary entropy~\cite{Affleck1991a}, as well as algebraic finite-size corrections~\cite{WuXT2012}. The crosscap condition instead implements an orientation-reversing identification and produces a nonorientable geometry without a physical boundary. The corresponding universal constant is the Klein bottle entropy $\ln |\langle 0|\mathcal{C}\rangle|^2$~\cite{TuHH2017}, where $|\mathcal{C}\rangle$ and $|0\rangle$ denote the crosscap state and the CFT vacuum, respectively. It therefore probes universal information associated with orientation reversal that is not captured by conventional conformal boundaries. While finite-size corrections have been extensively studied for conventional boundary conditions, the finite-size corrections governing how the lattice crosscap overlap approaches its universal CFT value $\langle 0|\mathcal{C}\rangle$, and hence the corresponding corrections to the Klein bottle entropy, remain much less explored.

The 2D classical Ising model provides an ideal platform for examining these concepts. Its critical point is described by the $c=1/2$ Ising CFT~\cite{Belavin1984,Francesco-Book}, and its exact solvability permits rigorous analytical treatment~\cite{Onsager1944,Kaufman1949a,Kaufman1949b,Kac1952,YangCN1952b,Schultz1964,McCoy-Book}. Recent work~\cite{ZhangYS2026a} established an explicit lattice construction of crosscap states in the quantum transverse-field Ising chain and derived their field theoretical counterparts. A notable finding is that the crosscap overlap in the quantum transverse-field Ising chain at criticality is free of finite-size corrections~\cite{ZhangYS2026a,KimJC2024b}. This result raises a natural question: away from the anisotropic limit — where the 2D classical Ising model maps onto the 1D quantum Ising chain — what is the scaling form of the finite-size corrections to the crosscap overlap along the critical line?

In this work, we address this question by analyzing the crosscap overlap in the 2D classical Ising model along its critical self-dual line, using the exact solution. We find that finite-size corrections exhibit an \emph{exponential} decay as a function of system size, except at the anisotropic limit, where corrections vanish. This exponential scaling is quite unusual and stands in sharp contrast to the powerlaw finite-size corrections~\cite{Izmailian2001,WuXT2012} typically generated by irrelevant operators in conformal perturbation theory~\cite{Cardy1986a}. We derive an exact finite-size correction formula using contour integral techniques and trace the origin of the exponential decay to branch point singularities in the complex plane.

{\em Crosscap overlap} --- We consider the 2D classical Ising model on an $ M \times N$ square lattice with spin variables $s_{ij} = \pm 1$, where $i = 1,\ldots,M$ and $j=1,\ldots,N$ label the row and column indices, respectively. Throughout this work we impose periodic boundary conditions along the horizontal ($x$) direction and \emph{crosscap} boundary conditions along the vertical ($y$) direction; see Fig.~\ref{fig:Figure1}. More explicitly, the partition function is written as
\begin{align}
    Z = \sum_{\{s_{ij}\}} Z_{\mathrm{cross}}(\{s_{1,j}\})Z_{\mathrm{bulk}}(\{s_{ij}\}) Z_{\mathrm{cross}}(\{s_{M,j}\})\, ,
    \label{eq:crosscap-Z-1}
\end{align}
where $Z_{\mathrm{bulk}}$ contains the Boltzmann weights associated with bulk nearest-neighbor interactions,
\begin{align}
    Z_{\mathrm{bulk}}(\{s_{ij}\}) &= e^{K_x\sum_{i=1}^{M} \sum_{j=1}^{N} s_{ij}s_{i,j+1}} \nonumber  \\
    &\phantom{=} \; \times e^{K_y\sum_{i=1}^{M-1} \sum_{j=1}^N s_{ij}s_{i+1,j}}
\end{align}
with $K_x$ and $K_y$ denoting the horizontal and vertical couplings, respectively. The factors $Z_{\mathrm{cross}}$ encode crosscap boundary interactions,
\begin{align}
    Z_{\mathrm{cross}}(\{s_{l,j}\})
    = e^{K_y\sum_{j=1}^{N/2}s_{l,j}s_{l,j+N/2}}\, ,\quad l = 1,M\, .
    \label{eq:crosscap-interactions}
\end{align}

To formulate the partition function in the transfer-matrix formalism, we identify each Ising spin $|s\rangle$ ($s=\pm 1$) with the eigenbasis of the Pauli matrix $\sigma^x$, and introduce row-to-row transfer matrices~\cite{Schultz1964,PlischkeBergersen-Book}:
\begin{align}
    V_1 &= [2\sinh(2K_y)]^{N/2} e^{K^{*}_y \sum_{j=1}^{N} \sigma_j^z} \, ,\nonumber\\
    V_2 &= e^{K_x \sum_{j=1}^{N} \sigma_j^x \sigma_{j+1}^x}
    \label{eq:transfer-mat}
\end{align}
with $K^{*}_y=\frac{1}{2} \ln [\coth(K_y)]$.
Equivalently, relative to the more common convention in which the Ising spin is represented by $\sigma^z$, this choice amounts to a uniform spin rotation $\sigma^x\to\sigma^z$, $\sigma^z\to-\sigma^x$, which leaves the spectrum unchanged but gives the transfer matrix the form used below. The factor $Z_{\mathrm{cross}}$ can be rewritten as
\begin{align}
    Z_{\mathrm{cross}}(\{s_{l,j}\}) = \langle \{s_{l,j}\}|\sqrt{V_1}|\mathcal{C}_{\mathrm{latt}}\rangle\, ,
\end{align}
where the lattice crosscap state is defined by
\begin{align}
    |\mathcal{C}_{\mathrm{latt}}\rangle
    = \prod_{j=1}^{N/2} \left( |{+1}\rangle_{j} |+1\rangle_{j+N/2}  + |{-1}\rangle_{j} |-1 \rangle_{j+N/2} \right) \, .
\label{eq:crosscap-state}
\end{align}

The partition function then takes a compact form
\begin{align}
    Z
    &= \langle \mathcal{C}_{\mathrm{latt}}| \sqrt{V_1} (V_2 V_1)^{M-1} V_2 \sqrt{V_1} |\mathcal{C}_{\mathrm{latt}}\rangle\nonumber\\
    &\equiv \langle \mathcal{C}_{\mathrm{latt}}| V^M|\mathcal{C}_{\mathrm{latt}}\rangle\, ,
    \label{eq:crosscap-Z-2}
\end{align}
where $V = \sqrt{V_1}V_2 \sqrt{V_1}$ is the transfer matrix of the Ising model.
In the limit $M\gg N$, only the leading eigenvalue and its associated eigenvector of the transfer matrix $V$ contribute in Eq.~\eqref{eq:crosscap-Z-2}, yielding the asymptotic expansion
\begin{align}
    \ln Z \simeq M\ln\Lambda_0 + \ln |\langle \psi_0|\mathcal{C}_{\mathrm{latt}}\rangle|^2 + \cdots\, ,
    \label{eq:logZ-asym-expan}
\end{align}
where $\Lambda_0$ is the leading eigenvalue of $V$, and $|\psi_0\rangle$ is the corresponding eigenvector. The first term represents the bulk free energy, while the second term, expressed as the logarithm of the squared norm of the \emph{crosscap overlap}, is known as the Klein bottle entropy~\cite{TuHH2017}.

The 2D classical Ising model is critical on the self-dual line $K_x=K_y^*$~\cite{Schultz1964,Kramers1941}. We will focus on this case in the present work. Its continuum limit is described by the Ising CFT. Then, the lattice crosscap overlap in Eq.~\eqref{eq:logZ-asym-expan} is expected to approach a universal constant in the thermodynamic limit,
\begin{align}
    \lim_{N\to\infty}\langle \psi_0 |\mathcal{C}_{\mathrm{latt}}\rangle = \sqrt{\frac{2+\sqrt{2}}{2}}\, ,
    \label{eq:cross-ovlp-Ising-CFT}
\end{align}
a value determined by the quantum dimensions of the Ising CFT~\cite{TuHH2017}. The purpose of this work is to derive the finite-size corrections to this universal value.

\begin{figure}[ht]
    \centering
    \includegraphics[width=1.0\linewidth]{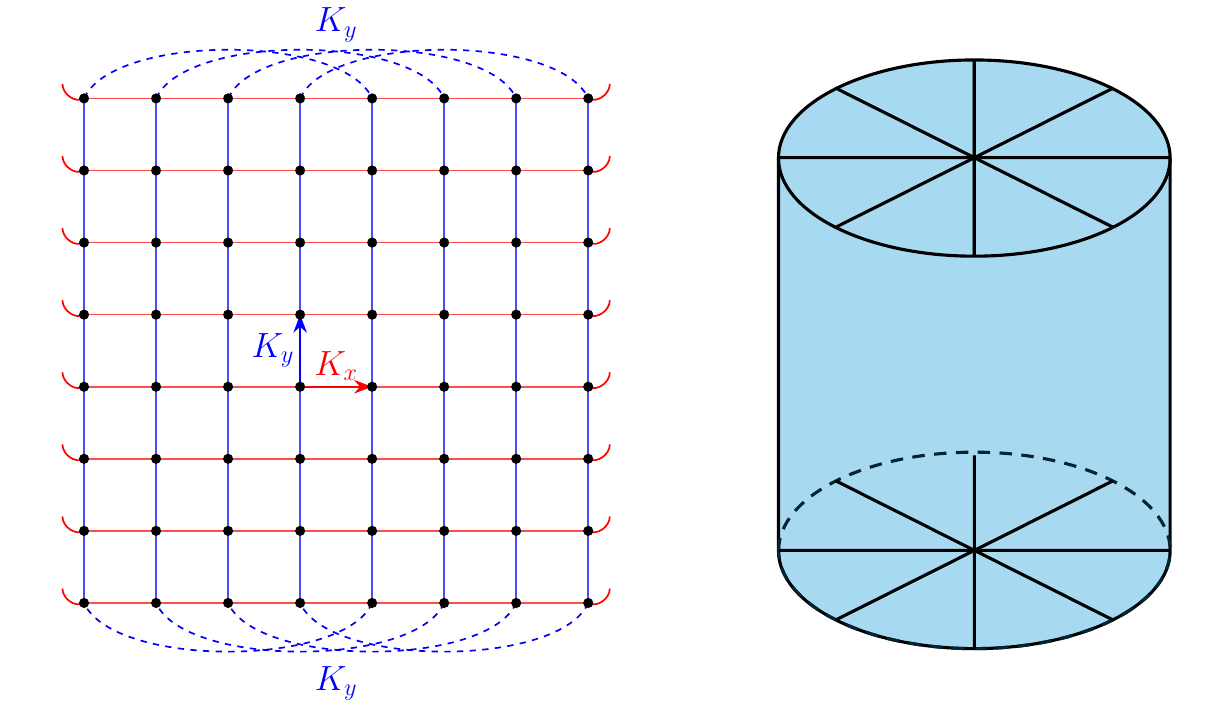}
    \caption{Left: schematic of the 2D classical Ising model with anisotropic couplings and periodic (crosscap) boundary conditions along the $x$ ($y$) direction. Right: geometric depiction of the cylinder with crosscap boundaries. }
    \label{fig:Figure1}
\end{figure}

Before proceeding to the derivation, it is worth noting that the transfer matrix $V$ in Eq.~\eqref{eq:crosscap-Z-2} commutes with the Hamiltonian of an anisotropic spin-$1/2$ quantum XY chain~\cite{Suzuki1971,Rams2015},
\begin{align}
    H = -\sum_{j=1}^N \left(\frac{1+\gamma}{2}\sigma^x_j\sigma^x_{j+1} + \frac{1-\gamma}{2}\sigma^y_j\sigma^y_{j+1} + h \sigma^z_j\right)\, ,
    \label{eq:XY-ham}
\end{align}
where the parameters $\gamma$ and $h$ are related to the Ising couplings via $\gamma = 1/\cosh (2K_y^*)$ and $h= \tanh (2K_y^*)/\tanh (2K_x)$. The transfer matrix $V$ and the Hamiltonian $H$ can be simultaneously diagonalized~\cite{Schultz1964,Suzuki1971}, and the leading eigenvector $|\psi_0\rangle$ of $V$ is the ground state of $H$~\cite{Suzuki1971}. Thus, they share the same crosscap overlap results, even for finite-size cases.

The transfer matrix $V$, as well as the Hamiltonian~\eqref{eq:XY-ham}, possesses a global $\mathbb{Z}_2$ symmetry, $[V,Q] = 0$, with $Q=\prod_{j=1}^N\sigma^z_j$. Both the leading eigenvector $|\psi_0\rangle$ and the lattice crosscap state $|\mathcal{C}_{\mathrm{latt}}\rangle$ [Eq.~\eqref{eq:crosscap-state}] live in the $\mathbb{Z}_2$ even ($Q=1$) sector, which, following the CFT convention, is called the Neveu-Schwarz (NS) sector. Utilizing the Jordan-Wigner transformation, $\sigma_{j}^z = 2c_{j}^\dagger c_{j} - 1$ and $\sigma_{j}^x = (c_{j}^\dagger + c_{j}) e^{i\pi\sum_{l=1}^{j-1} c_{l}^\dagger c_{l}}$, the Fourier transform, $c_j^\dag = \frac{1}{\sqrt{N}}\sum_{k\in\mathrm{NS}} e^{-ikj}c^\dag_k$ ($k\in\mathrm{NS}$ denotes allowed lattice momenta in the NS sector, $k = \pm[\pi -\frac{2\pi}{N}(n_{k}-\frac{1}{2})]$ with $n_k=1,2,\ldots,N/2$), and a Bogoliubov transformation, the transfer matrix $V$ in the NS sector can be diagonalized~\cite{Schultz1964,Pfeuty1970}. Along the critical line $K_x=K_y^*$, the leading eigenvector $\ket{\psi_0}$ reads
\begin{equation}
     \ket{\psi_0}
     =\prod_{k>0}\left[\sin{\frac{\theta (k)}{2}+i\cos{\frac{\theta (k)}{2}}}c_{k}^\dagger c_{-k}^\dagger\right]\ket{0}_c \, ,
\label{eq:psi0}
\end{equation}
where $k>0$ denotes positive momenta in the NS sector, $\theta(k)$ is the Bogoliubov angle:
\begin{equation}
    \theta(k)=\arctan \left[\frac{\tan(k/2)}{\cosh(2K_x)}\right] \, ,
\label{eq:Bogoliubov-phases}
\end{equation}
and $|0\rangle_c$ is the vacuum of the Jordan-Wigner fermions. Using the results in Ref.~\cite{ZhangYS2026a}, the lattice crosscap state in Eq.~\eqref{eq:crosscap-state} is rewritten as a superposition of two fermionic Gaussian states in the convention of Ref.~\cite{ZhangYS2026a}, which is recovered by rewriting Eq.~\eqref{eq:crosscap-state} in the $\sigma^z$ eigenbasis:
\begin{align}
|\mathcal{C}_{\mathrm{latt}}\rangle &= \frac{1-i}{2} \prod_{j=1}^{N/2} (1 + i c_j^\dagger c_{j+N/2}^\dagger)|0\rangle_c \notag\\
&+ \frac{1+i}{2} \prod_{j=1}^{N/2} (1 - i c_j^\dagger c_{j+N/2}^\dagger)|0\rangle_c.
\end{align}
The lattice crosscap overlap of the critical 2D classical Ising model can hence be calculated~\cite{ZhangYS2026a}, yielding
\begin{align}
    \langle \psi_0 |\mathcal{C}_{\mathrm{latt}}\rangle = \sqrt{2}\cos\Theta
\label{eq:analytical-result-of-overlap}
\end{align}
with $\Theta = \frac{\pi}{4} + \frac{1}{2} \sum_{k>0} (-1)^{n_k} \theta (k)$.

In the thermodynamic limit, the alternating sum in $\Theta$ becomes an integral and can be analytically calculated,
\begin{align}
    \lim_{N\to\infty}\sum_{k>0} (-1)^{n_k} \theta (k) = -\frac{1}{2}\int^{\pi}_0 \theta'(k)\, \mathrm{d}k = -\frac{\pi}{4}\, ,
\end{align}
where we used $\theta (k + \frac{2\pi}{N})-\theta (k) \approx \frac{1}{2}\theta'(k) \mathrm{d}k$ with $\mathrm{d}k = \frac{4\pi}{N}$. This leads to $\lim_{N\rightarrow \infty} \langle \psi_0 |\mathcal{C}_{\mathrm{latt}}\rangle = \sqrt{2} \cos \frac{\pi}{8}$, which agrees with the CFT prediction in Eq.~\eqref{eq:cross-ovlp-Ising-CFT}.

{\em Finite-size corrections} --- We now analyze the finite-size corrections to the crosscap overlap using Eq.~\eqref{eq:analytical-result-of-overlap}. We define the deviation from the CFT prediction as
\begin{align}
   \Delta := \left| \langle\psi_0|\mathcal{C}_{\mathrm{latt}}\rangle - \sqrt{\frac{2+\sqrt{2}}{2}} \right| \, .
\label{eq:crosscap-deviation}
\end{align}
Figure~\ref{fig:Figure2}(a) shows $\ln \Delta$ as a function of the system size $N$ for several Ising couplings $K_x$ along the self-dual line $(K_x=K_y^*)$. We find that the finite-size corrections decay \emph{exponentially} with system size, $\Delta \sim e^{-\alpha N}$, where $\alpha$ is a coupling-dependent decay constant. Moreover, $\alpha$ increases and diverges as the coupling approaches the anisotropic limit $K_x\to 0$, indicating that the finite-size corrections vanish in this limit. Indeed, on the self-dual line $K_x=K_y^*$ [corresponding to $h=1$ in Eq.~\eqref{eq:XY-ham}], the anisotropic limit $ K_x \to 0$ yields $\gamma=1/\cosh(2K_y^*) \to 1$. The Hamiltonian $H$ therefore reduces to the critical transverse-field Ising chain, whose crosscap overlap is known to be free of finite-size corrections~\cite{ZhangYS2026a,KimJC2024b}.

\begin{figure}[ht]
    \centering
    \includegraphics[width=1\linewidth]{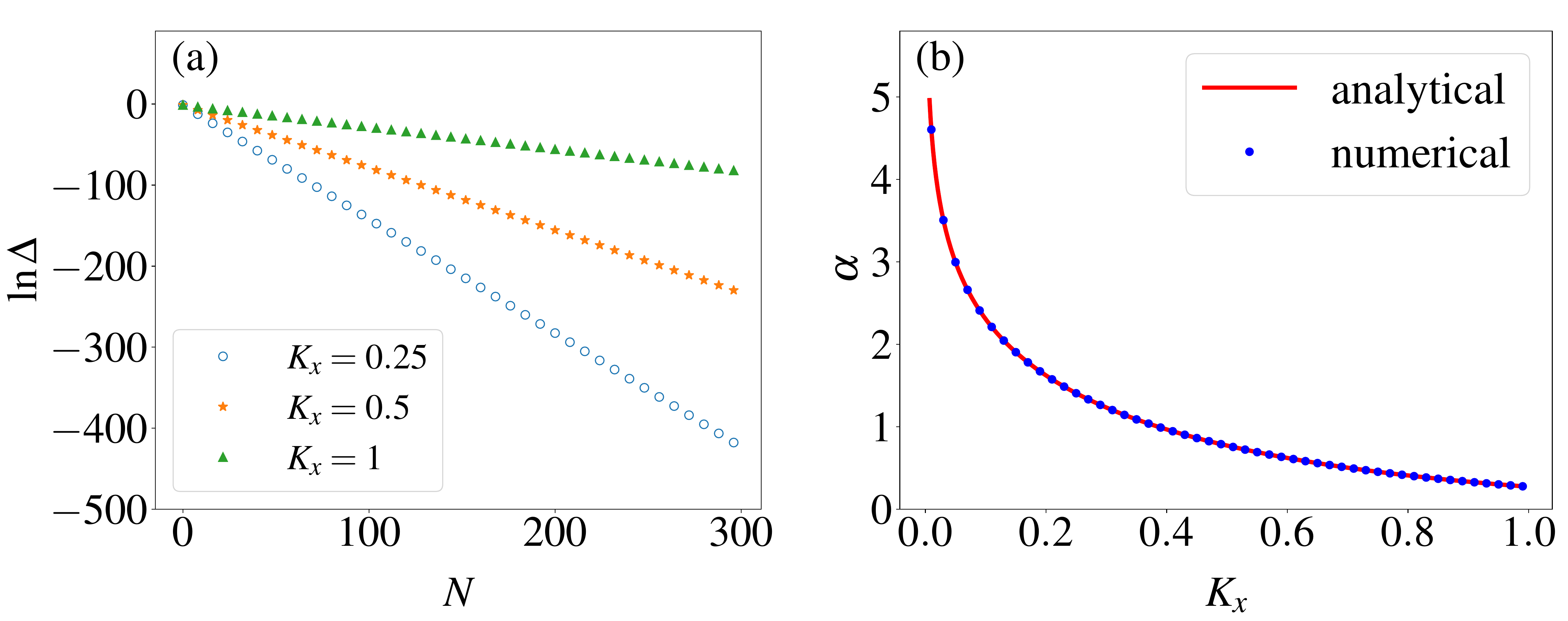}
    \caption{Finite-size corrections to the crosscap overlap for the 2D Ising model along the self-dual line ($K_x=K_y^*$). (a) Logarithm of the deviation $\ln \Delta$ [Eq.~\eqref{eq:crosscap-deviation}] as a function of the system size $N$ for several values of the coupling $K_x$. (b) Decay constant $\alpha$ as a function of $K_x$: numerical estimates (blue dots) extracted from the slopes in (a), compared with the analytical result (red solid line) in Eq.~\eqref{eq:decay-const}. }
\label{fig:Figure2}
\end{figure}

The exponential decay of the finite-size corrections implies that they are intrinsically non-perturbative, in the sense that they do not admit an expansion in powers of $1/N$. This motivates a non-perturbative analysis of the factor $\Theta$, which can be achieved by rewriting the alternating sum as a contour integral:
\begin{align}
    \sum_{k>0} (-1)^{n_k} \theta (k) =
    \frac{-1}{2\pi i} \oint_C \theta(z)\, \mathrm{d}\ln\varphi (z) \, ,
\label{eq:coutour-integral}
\end{align}
where $\varphi(z)=(e^{iNz/2}+i)/(e^{iNz/2}-i)$ is an auxiliary function. Its zeros are at $z^{(0)}_\ell=\frac{2\pi}{N}(2\ell-\frac{1}{2})$, while its poles are at $z^{(\infty)}_\ell=\frac{2\pi}{N}(2\ell+\frac{1}{2})$, with $\ell\in\mathbb{Z}$. Together, these points form the set $z_n=\frac{2\pi}{N}(n-\frac{1}{2})$, $n\in\mathbb{Z}$, which gives the logarithmic branch points of $\ln\varphi(z)$. Moreover, $\mathrm{d}\ln\varphi(z)=[\varphi'(z)/\varphi(z)]\mathrm{d}z$, with $\varphi'(z)/\varphi(z)=N e^{iNz/2}/(e^{iNz}+1)$, so Eq.~\eqref{eq:coutour-integral} follows by collecting the corresponding residues.

The contour $C$ encircles these branch points on the real axis for $n=1,\ldots,N/2$.

Since $\ln \varphi (z)$ is multi-valued, its value must be fixed consistently along $C$ by analytic continuation. To do so, we write $\varphi (z) = \varphi_+ (z)/\varphi_- (z)$ with $\varphi_\pm (z) := e^{iNz/2}\pm i$, and express each factor in polar form, $\varphi_\pm (z) = \rho_\pm (z) e^{i\phi_\pm (z)}$ with $\phi_\pm \in (-\pi,\pi)$. The local section of $\ln \varphi (z)$ along $C$ is then defined as
\begin{align}
    \ln \varphi (z) := \ln [\rho_+(z)/\rho_-(z)] + i [\phi_+(z) -\phi_-(z) + 2\pi m]\, ,
    \label{eq:lnphi}
\end{align}
where the integer $m\in\mathbb{Z}$ is the winding number that labels the section. The choice of $m$ fixes a local branch of $\ln\varphi(z)$ along the contour. The intermediate quantities $\ln(-1)$ and $\ln\varphi(z_0^+)$ below must be evaluated in the same section; changing $m$ shifts them consistently and leaves the final combination in Eq.~\eqref{eq:cont-full} unchanged. In the explicit evaluation below, we use the section $m=0$. With this prescription, $\ln \varphi (z)$ is single-valued along the closed contour $C$ when $\mathrm{mod}(N,4)=0$, as the phase contributions from the branch points associated with the zeros and poles of $\varphi (z)$ cancel exactly. In contrast, for $\mathrm{mod}(N,4)=2$, traversing the contour accumulates an additional phase shift of $2\pi$.

We first focus on the case $\mathrm{mod}(N,4)=0$, and the case $\mathrm{mod}(N,4)=2$ will be discussed afterwards. For $\mathrm{mod}(N,4)=0$, the monodromy of $\ln \varphi (z)$ allows integration by parts without boundary contributions, yielding
\begin{align}
    \frac{-1}{2\pi i} \oint_{C} \theta (z) \mathrm{d}\ln \varphi (z)
    = \frac{1}{2\pi i} \oint_{C} \theta' (z) \ln \varphi (z) \mathrm{d}z \, .
    \label{eq:int-by-parts}
\end{align}
For practical evaluation, we take the contour to be the boundary of the infinite strip $\{z: 0\leq\mathrm{Re}(z)\leq \pi\}$, deformed to pass through the poles of $\theta'(z)$ at $z^\pm_0 = \pi \pm 2i\alpha$,
with
\begin{align}
    \alpha = \frac{1}{2}\ln \left[\frac{\cosh (2K_x)+1}{\cosh (2K_x)-1}\right]\, ,
    \label{eq:decay-const}
\end{align}
as illustrated in Fig.~\ref{fig:Figure3}. We will show that the constant $\alpha$ defined in Eq.~\eqref{eq:decay-const} indeed governs the exponential decay of the finite-size corrections, in agreement with the numerical results [see Fig.~\ref{fig:Figure2}(b)].

\begin{figure}[ht]
    \centering
    \includegraphics[width=0.8\linewidth]{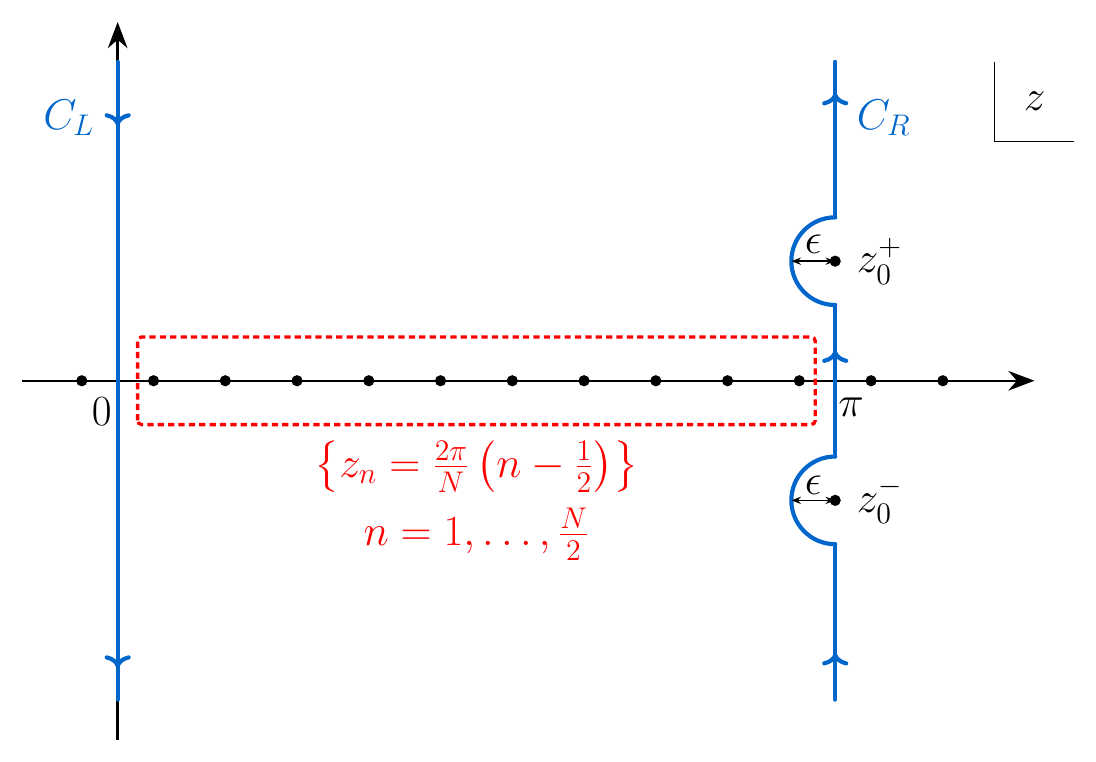}
    \caption{Contour integration for analyzing finite-size corrections. The contour $C$ is chosen to be a rectangle path extending infinitely along the imaginary axis, with small arcs around the branch points. It encloses the real-axis singularities of $\mathrm{d}\ln\varphi(z)$ corresponding to the allowed $k>0$ momenta in the NS sector.}
    \label{fig:Figure3}
\end{figure}

The contour can be divided into the left and right parts, $C=C_L + C_R$, with the right part further split as $C_R = \tilde{C}_R + C^+_\epsilon + C^-_\epsilon$. The integrals over these segments can then be evaluated term by term. The contributions from $C_\epsilon^\pm$ are computed via the residue theorem,
\begin{align}
    \frac{1}{2\pi i} \int_{C_\epsilon^\pm}
    \theta'(z) \ln\varphi (z) \mathrm{d}z
    &= -\frac{1}{2}\mathrm{Res}[\theta'(z^\pm_0)] \ln \varphi(z^\pm_0) \, ,
    \label{eq:cont1}
\end{align}
where $\mathrm{Res}[\theta'(z^\pm_0)] = \mp \frac{i}{2}$ are the residues of $\theta' (z)$ at $z^\pm_0$.

The remaining integrals over $\tilde{C}_R$ and $C_L$ can be calculated using the symmetry properties of $\theta'(z)$ and $\varphi (z)$,
\begin{align}
    \theta'(z) = \theta'(\bar{z})\, ,\quad
    \varphi (z) \varphi (\bar{z}) = -1\, ,
    \label{eq:symm}
\end{align}
for all $z=iy$ and $z=\pi + iy$ with $y\in\mathbb{R}$. For instance, after the substitution, $z = \pi + i y$, the integral along $\tilde{C}_R$ becomes
\begin{align}
    &\phantom{=} \; \frac{1}{2\pi i} \int_{\tilde{C}_R} \theta'(z) \ln \varphi (z) \mathrm{d}z\nonumber\\
    &= \frac{\ln (-1)}{2\pi} \left(\int_0^{2\alpha-\epsilon} + \int_{2\alpha+\epsilon}^\infty\right)  \,\theta'(\pi + iy) \mathrm{d}y\, ,
    \label{eq:cont2-intermed}
\end{align}
where
\begin{align}
    \ln (-1) := \ln [\varphi(iy)\varphi(-iy)] =\ln [\varphi(\pi+iy)\varphi(\pi-iy)]
    \label{eq:log-1}
\end{align}
is a well-defined constant for all $y\in \mathbb{R}$ due to its continuity along the contour: $\{z: \mathrm{Re}(z)=0\}\cup \{z: \mathrm{Re}(z)=\pi\}$. This can be re-expressed as a contour integral,
\begin{align}
    \text{Eq.}~\eqref{eq:cont2-intermed}
    &= \frac{\ln (-1)}{2\pi i} \int_{\tilde{C}_R^+} \theta'(z) \mathrm{d}z\nonumber\\
    &= \frac{\ln (-1)}{2\pi i} \int_{C_R^+} \theta'(z) \mathrm{d}z  + \frac{\ln (-1)}{2} \mathrm{Res}[\theta'(z^+_0)]\, ,
    \label{eq:cont2}
\end{align}
where $\tilde{C}_R^+$ denotes the upper half of $\tilde{C}_R$, i.e., $C_R^+ = \tilde{C}_R^+ + C_\epsilon^+$, and the residue theorem has been applied. Similarly, for the left contour $C_L$, we have
\begin{align}
    \frac{1}{2\pi i} \int_{C_L} \theta'(z) \ln \varphi (z) \mathrm{d}z
    = \frac{\ln (-1)}{2\pi i} \int_{C^+_L} \theta'(z) \mathrm{d}z\, ,
    \label{eq:cont3}
\end{align}
where $C_L^+$ is the upper half part of $C_L$, and $\ln (-1)$ has been defined in Eq.~\eqref{eq:log-1}.

Combining the results in Eqs.~\eqref{eq:cont1}, ~\eqref{eq:cont2}, and \eqref{eq:cont3}, the full contour integral over $C$ becomes
\begin{align}
    &\phantom{=} \;\frac{1}{2\pi i}\oint_C \theta'(z) \ln \varphi (z) \mathrm{d}z\nonumber\\
    &= \frac{\ln (-1)}{2\pi i} \int_{C^+} \theta'(z) \mathrm{d}z -\frac{i}{2} \left[\ln (-1) - \ln \varphi(z^+_0)\right]\, ,
    \label{eq:cont-full}
\end{align}
where $C^+ = C_L^+ + C_R^+$, and the associated integral evaluates to $\int_{C^+} \theta'(z) \mathrm{d}z = \theta (0) - \theta (\pi) = -\frac{\pi}{2}$. Although the final result in Eq.~\eqref{eq:cont-full} is independent of the section chosen for $\ln \varphi (z)$, the explicit values of $\ln (-1)$ and $\ln \varphi (z^+_0)$ do depend on this choice. To obtain a consistent expression, both must be computed in the same section. In a specific section, e.g., $m=0$ in Eq.~\eqref{eq:lnphi}, we obtain $\ln (-1) = i\pi$ and $\ln \varphi(z_0^+) = i(\pi - 2\delta)$, with $\delta = \arctan (e^{-\alpha N})$. Substituting these expressions into Eq.~\eqref{eq:cont-full} yields
\begin{align}
    \sum_{k>0} (-1)^{n_k} \theta (k)
    = \frac{1}{2\pi i}\oint_C \theta'(z) \ln \varphi (z) \mathrm{d}z
    = -\frac{\pi}{4} + \delta\, .
\label{eq:finite-size-corr}
\end{align}

The analysis for $\mathrm{mod}(N,4)=2$ proceeds similarly, although it is slightly more involved: an extra boundary term appears in the integration-by-parts step [Eq.~\eqref{eq:int-by-parts}], and the value of $\ln (-1)$ differs along the left and right contours due to the $2\pi$ phase shift accumulated around the full contour. Nevertheless, the final expression remains the same as in Eq.~\eqref{eq:finite-size-corr}.

Equation \eqref{eq:finite-size-corr} is the key result of this paper. Substituting it into Eq.~\eqref{eq:analytical-result-of-overlap} yields a remarkably simple formula for the lattice crosscap overlap: $\langle\psi_0|\mathcal{C}_{\mathrm{latt}}\rangle = \sqrt{2}\cos (\frac{\pi}{8}+\frac{1}{2}\delta)$. For $N\gg 1$, we have $\delta \simeq e^{-\alpha N}$, and the crosscap overlap admits the asymptotic expansion
\begin{align}
    \langle\psi_0|\mathcal{C}_{\mathrm{latt}}\rangle
    = \sqrt{\frac{2+\sqrt{2}}{2}} - \frac{1}{2} \sqrt{\frac{2-\sqrt{2}}{2}}\, e^{-\alpha N} + \cdots\, ,
    \label{eq:final-result}
\end{align}
which exactly reproduces the exponentially decaying finite-size corrections observed in Fig.~\ref{fig:Figure2}(a). The decay constant $\alpha$, given in Eq.~\eqref{eq:decay-const}, depends on the Ising coupling and encodes the singularity of the Bogoliubov angle in the complex plane.
In fact, the complex singularity of the Bogoliubov angle is directly related to the Toeplitz determinant kernel of the Ising model under the variable transformation $z'=e^{i(\pi-k)}$~\cite{WuTT1966,McCoy-Book,Deift2013}. This connection provides a more transparent derivation of Eq.~\eqref{eq:final-result}, in which the exponential finite-size correction is directly associated with the square-root branch-cut structure of the Toeplitz kernel. It also provides a natural framework for extending the analysis to the off-critical region. See the Supplemental Material~\cite{Append} for further details~\footnote{The authors thank the anonymous referee for explicitly pointing out these observations.}.

Finally, we note that the overlaps of the crosscap state with all eigenvectors of the transfer matrix $V$ can be calculated analytically. Actually, only eigenvectors of the form $|\psi_{k_1\cdots k_P}\rangle = \prod_{l=1}^P d^\dag_{k_l} d^\dag_{-k_l}|\psi_0\rangle$ have nonvanishing overlaps with $|\mathcal{C}_{\mathrm{latt}}\rangle$, with $d_k = \sin [\theta (k)/2] c_k - i\cos [\theta (k)/2] c^\dag_{-k}$ being the Bogoliubov quasiparticle operators annihilating $|\psi_0\rangle$ [Eq.~\eqref{eq:psi0}]. These excited-state crosscap overlaps are given by
\begin{align}
    |\langle \psi_{k_1\cdots k_P}|\mathcal{C}_{\mathrm{latt}}\rangle| =
    \begin{cases}
        \sqrt{2}\cos \Theta \, , & P \; \mathrm{even}\,  \\
        \sqrt{2}\sin \Theta \, , & P \; \mathrm{odd}\,
    \end{cases},
\end{align}
where $\Theta$ is defined below Eq.~\eqref{eq:analytical-result-of-overlap}. This immediately shows that all nonvanishing crosscap overlaps in the 2D critical Ising model exhibit exponentially decaying finite-size corrections, with the same decay constant $\alpha$ given in Eq.~\eqref{eq:decay-const}.

{\em Summary and outlook} --- To summarize, we have analyzed the finite-size corrections to the crosscap overlap in the 2D classical Ising model along its self-dual critical line. We have demonstrated that the corrections decay exponentially with system size, originating from the branch-point singularities of the Bogoliubov angle treated as a complex function of lattice momentum.

An important open question is whether such exponentially decaying finite-size corrections are a special feature of the 2D Ising model, or arise more generally in other critical systems. While exponential suppression is highly advantageous for numerically locating critical points using crosscap overlaps, powerlaw finite-size corrections are known to occur in other exactly solvable models, such as the spin-1/2 Haldane-Shastry chain~\cite{TanBY2025,Haldane1988a,Shastry1988}. Future work should further explore the finite-size scaling of crosscap overlaps in more general microscopic models at criticality, in order to clarify under what conditions exponential versus algebraic behavior emerges. We leave this to future work.

{\em Acknowledgments} ---  We thank both anonymous referees for many helpful comments and suggestions. Y.T.Z is very grateful to Professor Chenjie Wang for guidance in related studies. He also thanks Haochen Tu and Tianyuan Zhou for illuminating discussions. Y.T.Z. acknowledges the financial support provided by The University of Hong Kong for funding his research internship at LMU Munich, where this work was initiated. He also sincerely thanks Professor Jan von Delft for the hospitality and support during the internship. H.-H.T. and Y.S.Z. are grateful to Lei Wang and Ying-Hai Wu for collaborations on related topics. L.P.Y. is supported by the National Natural Science Foundation of China (Grant No. 11874095). Y.S.Z. is supported by the Sino-German (CSC-DAAD) Postdoc Scholarship Program and Research Grants Council of Hong Kong (GRF 17311322, CRF C7015-24GF, CRS HKU701/24, and AoE/P-604/25-R).

\bibliographystyle{apsrev4-1}
\bibliography{refs}

\clearpage

\appendix

\begin{widetext}

\begin{center}
\textbf{Supplemental Material for ``Finite-size corrections to the crosscap overlap in the two-dimensional Ising model''}
\end{center}

\setcounter{table}{0}
\renewcommand{\thetable}{S\arabic{table}}
\setcounter{figure}{0}
\renewcommand{\thefigure}{S\arabic{figure}}
\setcounter{equation}{0}
\renewcommand{\theequation}{S\arabic{equation}}

This Supplemental Material extends the lattice calculation of the crosscap overlap presented in the main text to the off-critical Ising model. We relate the Bogoliubov-angle formulation to the standard Toeplitz determinant kernel (Sec.~I A)
and derive an exact contour-integral representation of the lattice crosscap overlap valid both at and away from criticality (Sec.~I B). The scaling limit of the lattice result then reproduces the universal scaling function of the crosscap overlap for the Ising conformal field theory with thermal perturbation.

\maketitle
\tableofcontents

\section{I. Lattice crosscap overlap in the off-critical region}

\subsection{A. Bogoliubov angle and the Toeplitz determinant kernel}

The row-to-row transfer matrix $V$ of the classical Ising model shares the same eigenvectors as the anisotropic spin-$1/2$ quantum XY chain Hamiltonian
[Eq.~(10) of the main text],
\begin{align}
    H = -\sum_{j=1}^N \left(\frac{1+\gamma}{2}\sigma^x_j\sigma^x_{j+1} + \frac{1-\gamma}{2}\sigma^y_j\sigma^y_{j+1} + h \sigma^z_j\right)\, ,
    \label{eq:XY-ham-SM}
\end{align}
where the parameters $\gamma$ and $h$ are related to the Ising couplings $K_x$ and $K_y$ via
\begin{align}
    \gamma = 1/\cosh (2K_y^*)\, ,\quad h= \tanh (2K_y^*)/\tanh (2K_x)\, ,\label{eq:XY-ham-paramters-SM}
\end{align}
with $K_y^* = \frac{1}{2}\ln [\coth(K_y)]$. The model is critical for $h=1$, and is in the ferromagnetically ordered (disordered) phase for $h<1$ ($h>1$).

The Hamiltonian $H$ and the Ising transfer matrix $V$ can be diagonalized
simultaneously by the same Jordan-Wigner and Bogoliubov transformations as
those used in the main text. The corresponding Bogoliubov angle is
\begin{equation}
    \tan\theta(k)
    =
    \frac{\sinh(2K_x)\sin k}
    {\cosh(2K_x)\sinh(2K_y^*)
    +\sinh(2K_x)\cosh(2K_y^*)\cos k}\, ,
    \label{eq:theta-general-SM}
\end{equation}
which reduces to Eq.~(12) of the main text on the self-dual line $K_x=K_y^*$.

Since the derivation of the lattice crosscap overlap relies only on the
Gaussian structure of the ground state $|\psi_0\rangle$ and the lattice
crosscap state $|\mathcal{C}_{\mathrm{latt}}\rangle$, the overlap formula
remains valid away from criticality:
\begin{equation}
    \langle\psi_0|\mathcal{C}_{\mathrm{latt}}\rangle
    =
    \sqrt{2}\cos\Theta,\qquad
    \Theta=
    \frac{\pi}{4}
    +
    \frac{1}{2}
    \sum_{n=1}^{N/2}(-1)^n\theta(k_n),
    \label{eq:theta-sum-SM}
\end{equation}
where
\begin{equation}
    k_n=\pi-\frac{2\pi}{N}\left(n-\frac{1}{2}\right),
    \qquad n=1,\ldots,N ,
\end{equation}
are the allowed lattice momenta in the NS sector.

To make contact with the standard Toeplitz-symbol description of the Ising model, it is convenient to introduce
\begin{equation}
    \lambda_1 :=e^{-2K_y}\coth K_x,\qquad
    \lambda_2 :=e^{-2K_y}\tanh K_x ,
    \label{eq:delta-def-SM}
\end{equation}
in terms of which Eq.~\eqref{eq:theta-general-SM} becomes
\begin{align}
    \tan\theta (k) &= \frac{\sinh (2K_y)\sin k}{\coth (2K_x) + \cosh (2K_y)\cos k}\nonumber\\
    &= \frac{(e^{2K_y}-e^{-2K_y})\sin k}{\coth(K_x)+\tanh(K_x) + (e^{2K_y}+ e^{-2K_y})\cos k}\nonumber\\
    &= \frac{(1-\lambda_1\lambda_2)\sin k}
    {(\lambda_1 +\lambda_2)+(1+\lambda_1\lambda_2)\cos k}\, ,
    \label{eq:theta-general-2-SM}
\end{align}
where we used the identities
\begin{align}
    \cosh(2K_y^*) = \coth (2K_y)\, ,\quad \sinh(2K_y^*) = 1/\sinh(2K_y)\, .
\end{align}
From
\begin{align}
    h = \frac{1}{\tanh (2K_x)\cosh (2K_y)}
    = \frac{\coth (K_x) + \tanh (K_x)}{e^{2K_y} + e^{-2K_y} }
    \, ,
\end{align}
one finds
\begin{align}
   0<\lambda_2 < \lambda_1 <1
\end{align}
in the ordered region, while
\begin{align}
    0<\lambda_2 < \lambda_1^{-1} <1
\end{align}
in the disordered region.

The key observation is that the Bogoliubov angle can be expressed directly
in terms of the standard Toeplitz determinant kernel for the Ising model~\cite{WuTT1966,McCoy-Book,Deift2013}
\begin{align}
    K_{\lambda} (z) := \sqrt{\frac{1-\lambda z}{1-\lambda z^{-1}}}\, ,\quad 0<\lambda <1\, ,
    \label{eq:Toeplitz-kernel-SM}
\end{align}
whose branch cuts are located at $(0,\lambda)\cup (\lambda^{-1},\infty)$.

Indeed, Eq.~\eqref{eq:theta-general-2-SM} implies
\begin{align}
    e^{2i\theta (k)} = \frac{1+i\tan\theta (k)}{1-i\tan\theta (k)}= \frac{(\lambda_1 + \lambda_2) + e^{ik} + \lambda_1\lambda_2 e^{-ik}}{(\lambda_1 + \lambda_2) + e^{-ik} + \lambda_1\lambda_2 e^{ik}}= \frac{(e^{ik} + \lambda_1)(1+ \lambda_2 e^{-ik})}{(e^{-ik} + \lambda_1)(1+ \lambda_2 e^{ik})}\, .
\end{align}
Let
\begin{align}
    z = e^{i(\pi-k)}\, ,
\end{align}
one obtains
\begin{align}
    e^{i\theta (k)} = e^{ik} K_{\lambda_1}(z) K_{\lambda_2} (z) = [K_{\lambda_1^{-1}}(z)]^{-1} K_{\lambda_2} (z)\, .
\end{align}
Therefore,
\begin{align}
    \theta (k) -k = -i\left[\ln K_{\lambda_1}(z) + \ln K_{\lambda_2} (z)\right]\, ,
    \label{eq:theta-ordered-SM}
\end{align}
in the ordered region, whereas
\begin{align}
    \theta (k)  = -i\left[\ln K_{\lambda_2}(z) - \ln K_{\lambda_1^{-1}} (z)\right]\, ,
    \label{eq:theta-disordered-SM}
\end{align}
in the disordered region.

Equations~\eqref{eq:theta-ordered-SM} and
\eqref{eq:theta-disordered-SM} establish a direct connection between the crosscap overlap and the analytic structure of the Ising Toeplitz kernel.
As shown below, the lattice crosscap overlap is governed by the square-root branch
singularities of $K_{\lambda}(z)$ in the complex $z$-plane.

\subsection{B. Evaluation of the crosscap overlap from contour integrals}

Using the oddness of the Bogoliubov angle, $\theta (-k) = -\theta (k) $, the alternating sum appearing in Eq.~\eqref{eq:theta-sum-SM} can be extended
to the full NS momentum sector:
\begin{align}
    \sum_{n=1}^{N/2} (-1)^n \theta (k_n) = \frac{1}{2}\sum_{n=1}^N (-1)^n \theta (k_n)\, .
\end{align}
Furthermore,
\begin{align}
    \sum_{n=1}^N (-1)^n k_n = -\pi\, ,
\end{align}
so that
\begin{align}
    \Theta = \frac{\pi}{4} + \frac{1}{4}\sum_{n=1}^N (-1)^n \theta (k_n)
    = \frac{1}{4}\sum_{n=1}^N (-1)^n [\theta (k_n) - k_n]\, .
\end{align}
Substituting Eqs.~\eqref{eq:theta-ordered-SM} and
\eqref{eq:theta-disordered-SM}, the phase $\Theta$ can be expressed entirely in terms of logarithms of Toeplitz kernels:
\begin{align}
    \Theta = -\frac{i}{4}\sum_{n=1}^N (-1)^n \ln K_{\lambda_1} (z_n) -\frac{i}{4} \sum_{n=1}^N (-1)^n \ln K_{\lambda_2} (z_n)
\end{align}
in the ordered region, while
\begin{align}
    \Theta = \frac{\pi}{4} + \frac{i}{4}\sum_{n=1}^N (-1)^n \ln K_{\lambda_1^{-1}} (z_n) -\frac{i}{4} \sum_{n=1}^N (-1)^n \ln K_{\lambda_2} (z_n)
\end{align}
in the disordered region, where
\begin{align}
    z_n := e^{i(\pi-k_n)} = \exp\left[{i\frac{2\pi}{N}}\left(n-\frac{1}{2}\right)\right]\, ,\quad n=1,\ldots,N\, .
\end{align}

It is therefore sufficient to
evaluate the alternating sum
\begin{align}
    S_\lambda
    :=
    -\frac{i}{4}
    \sum_{n=1}^{N}
    (-1)^n
    \ln K_\lambda(z_n).
\end{align}
The points $\{z_n\}$ are precisely the zeros and poles of
$(z^{N/2}+i)/(z^{N/2}-i)$, which allows the sum to be converted into a contour integral:
\begin{align}
    S_{\lambda}
    &= -\frac{i}{4} \frac{1}{2\pi i}\oint_C \ln K_{\lambda}(z) \, \mathrm{d} \ln \left(\frac{z^{N/2}+i}{z^{N/2}-i}\right) \nonumber\\
    &= -\frac{1}{4} \oint_C \frac{\mathrm{d}z}{2\pi i z} \frac{N z^{N/2}}{z^N +1} \ln K_{\lambda} (z)\, ,
    \label{eq:Sum-contour-SM}
\end{align}
where the contour $C$ is the boundary of a thin annulus enclosing the unit circle, see Fig.~\ref{fig:contour-SM}(a). The contour may now be deformed onto the branch cuts of $\ln K_\lambda(z)$ located at $(0,\lambda)\cup(\lambda^{-1},\infty)$, see Fig.~\ref{fig:contour-SM}(b).
Across either cut, the logarithm acquires the discontinuity
\begin{align}
    \ln K_{\lambda} (t - i 0^+) - \ln K_{\lambda} (t + i 0^+) = i\pi\, ,\quad \forall t\in (0,\lambda)\cup (\lambda^{-1},\infty)\, .
\end{align}
The contour integral therefore reduces to
\begin{align}
    S_\lambda = \frac{1}{4} \left[\int_{0}^\lambda \frac{\mathrm{d}t}{2t} \frac{N t^{N/2}}{t^N +1} + \int_{\lambda^{-1}}^\infty \frac{\mathrm{d}t}{2t} \frac{N t^{N/2}}{t^N +1}\right] \, .
\end{align}
Using the transformation $t\to t^{-1}$ in the second integral gives
\begin{align}
    S_\lambda
    = \frac{1}{4} \int_{0}^\lambda \frac{\mathrm{d}t}{t} \frac{N t^{N/2}}{t^N +1}
    = \frac{1}{2}\arctan \lambda^{N/2}\, .
\end{align}
This result makes the origin of the exponential finite-size correction
transparent: the contribution of a given Toeplitz factor is controlled by
the branch cuts along $(0,\lambda)\cup (\lambda^{-1},\infty)$, producing the characteristic factor $\lambda^{N/2}$.

\begin{figure}[t]
    \centering
    \includegraphics[width=1.0\linewidth]{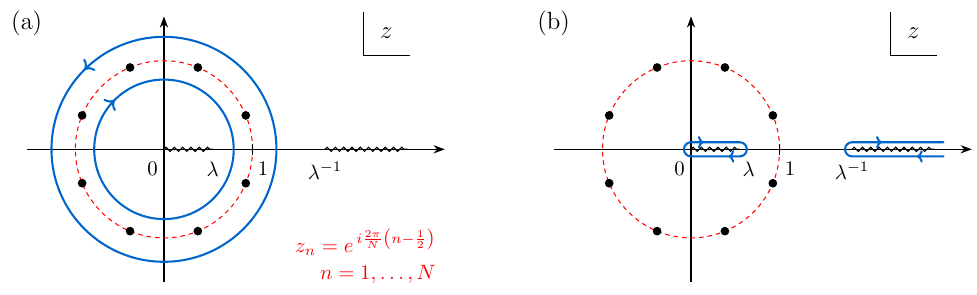}
    \caption{Contour integration in the $z$-plane.
    (a) The contour $C$ (blue) of the integral in Eq.~\eqref{eq:Sum-contour-SM} is taken to be the
    boundary of a thin annulus enclosing the unit circle. The branch cuts of $\ln K_\lambda(z)$, located along $(0,\lambda)\cup(\lambda^{-1},\infty)$, are indicated by the black zigzag lines. (b) The contour $C$ is deformed onto the branch cuts of $\ln K_\lambda(z)$, reducing the contour integral to integrals over the discontinuities across the cuts.}
    \label{fig:contour-SM}
\end{figure}

The phase $\Theta$ is therefore
\begin{align}
    \Theta = S_{\lambda_1} + S_{\lambda_2} = \frac{1}{2}\arctan \lambda_1^{N/2} + \frac{1}{2}\arctan \lambda_2^{N/2}
\end{align}
in the ordered region. In the disordered region,
\begin{align}
    \Theta &= \frac{\pi}{4} - S_{\lambda_1^{-1}} + S_{\lambda_2} \nonumber\\
    &= \frac{1}{2}\left(\frac{\pi}{2}  -\arctan \lambda_1^{-N/2} \right)+ \frac{1}{2}\arctan \lambda_2^{N/2}\nonumber\\
    &= \frac{1}{2}\arctan \lambda_1^{N/2} + \frac{1}{2}\arctan \lambda_2^{N/2}\, .
\end{align}
Hence both phases are described by the same final expression, and the lattice crosscap overlap becomes
\begin{align}
    \langle\psi_0|\mathcal{C}_{\mathrm{latt}}\rangle = \sqrt{2}\cos \left[\frac{1}{2}\arctan \lambda_1^{N/2} + \frac{1}{2}\arctan \lambda_2^{N/2}\right]\, .
    \label{eq:crossovlp-SM}
\end{align}

Equation~\eqref{eq:crossovlp-SM} allows us to extract the universal scaling
function of the crosscap overlap away from criticality. We consider the
scaling limit in which the Ising model approaches the critical point,
\begin{align}
    \lambda_1 :=e^{2\pi\epsilon}\rightarrow1,
\end{align}
while the system size is taken to infinity, $N\rightarrow\infty$. In this
limit, the infrared theory is described by the Ising conformal field theory
perturbed by the thermal operator, and the dimensionless scaling variable
\begin{align}
    s_1:=\epsilon N
\end{align}
is kept fixed, which plays the role of the effective coupling. One then finds
\begin{align}
    \lim_{\substack{\epsilon N=s_1\\N\rightarrow\infty}}
    \arctan\lambda_1^{N/2}
    =
    \arctan(e^{\pi s_1}),
    \qquad
    \lim_{N\rightarrow\infty}
    \arctan\lambda_2^{N/2}
    =
    0,
\end{align}
where the second limit follows from $0<\lambda_2<1$. Consequently, the lattice
crosscap overlap reduces to
\begin{align}
    \langle\psi_0(s_1)|\mathcal{C}_{\mathrm{latt}}\rangle
    =
    \sqrt{2}
    \cos\!\left[
        \frac12\arctan(e^{\pi s_1})
    \right]
    =
    \sqrt{
        1+\frac{1}{\sqrt{1+e^{2\pi s_1}}}
    },
\end{align}
which is in exact agreement with the universal scaling function obtained previously in
Refs.~\cite{ZhangYS2023,ZhangYS2026a}.

\end{widetext}

\end{document}